\documentclass[11pt]{article}
 \newcommand{\be}{\begin{equation}}
\newcommand{\ee}{\end{equation}}
 \newcommand{\beq}{\begin{eqnarray}}
\newcommand{\eeq}{\end{eqnarray}} 

\def\theequation{\arabic{section}.\arabic{equation}} 
\usepackage{graphicx}

\begin{document} 
 \begin{center} {\bf \LARGE Quantum Tachyon
Dynamics}\\ [7mm] H.  M.  FRIED
 \\ {\em Department of Physics \\ Brown University \\ Providence R.I.
02912 USA}\\ [5mm] Y.  GABELLINI
\\ {\em Institut Non Lin\'eaire de Nice\\ UMR 6618 CNRS\\ 1361 Route
des Lucioles\\ 06560 Valbonne France}\\ [5mm]

\vspace{5mm} Abstract
 \end{center}

It is suggested that charged tachyons of extremely large mass $M$ could not only
contribute to the dark matter needed to fit astrophysical observations, but
could also provide an explanation for gamma-ray bursts and ultra high energy
cosmic rays.  The present paper defines a quantum field theory of tachyons, of
particles similar to ordinary leptons, but with momenta larger than energy.  The
theory is invariant under the full $CPT$ transformation, but separately violates
$P$ and $T$ invariance.  Micro-causality is broken for space-time intervals
smaller than $M^{-1}$, but is effectively preserved for larger separations.
Charged fermionic, rather than charged scalar tachyons are considered in order
to minimize the probability of Cerenkov like radiation by the tachyon, thereby
permitting a high-energy tachyon to retain its energy over galactic distances.

Topics treated include the choice and Schwinger Action Principle variations of
an appropriate Lagrangian, spinorial wave functions, relevant Green's functions,
a functional description of an $S$--Matrix and generating functional, and a
variety of interesting kinematical processes, including photon emission and
reabsorption, and relevant annihilation and scattering effects.  A version of
Ehrenfest's theorem is developed, in order to provide a foundation for a
classical description of charged tachyons in an external elecromagnetic field.
Applications are then made to three outstanding astrophysical puzzles :  dark
matter, gamma ray bursts and ultra high energy cosmic rays.  
\newpage
{\bf\section{Introduction}} \setcounter{equation}{0} 
There are several
astrophysical puzzles of high current interest -- dark matter, the source and
content of the vacuum energy needed for an accelerating universe, the origins of
gamma ray bursts and ultra high energy cosmic rays coming from galactic
distances -- which may find a partial solution under the assumption that
massive, charged tachyons exist in, and in the outer reaches of, each galaxy.
The origin of tachyon pairs can be imagined in the immediate aftermath of the
Big Bang, or more locally, by any cataclysmic event, such as a supernova
explosion or black hole formation; they are definitely suggested by pair
production in the intense, electromagnetic vacuum field fluctuations appearing
in a `` bootstrap '' solution at extremely short distances \cite{one}.  In this
note, a brief QFT of tachyons as `` misgenerated '' leptons is presented, along
with a sketch of their interactions with themselves, with photons, and with
ordinary matter, in order to illustrate the possible relevance of such charged
tachyons to the experimental puzzles stated above.

Two immediate theoretical observations can be made to set the stage for this
discussion.  The Lagrangians typically used to define the Higgs mechanism, and
in particular the Weinberg -- Salam method of defining a `` true '' ground
state, are basically theories of ``~condensed '' tachyons, theories which use a
negative (mass)$^2$ and a self interaction to break spontaneously the initial
vacuum symmetry, and arrive at a minimum energy ground state.  But if a strong,
external, electromagnetic field is locally present, and if the quanta of the
tachyon field are charged, that ground state need not be, locally, the correct
vacuum state; and one might expect fluctuations about that W-- S vacuum.  Those
fluctuations could well be tachyonic, as expected from the form of the
Lagrangian used, suggesting that pairs of charged tachyons could, in principle
and just as for ordinary leptonic pairs, be produced by extremely strong
electromagnetic fields.

Secondly, consider the non perturbative Schwinger mechanism \cite{two} for the
production of lepton pairs by a strong electric field, a calculation which has
received during the past half century a great amount of attention and
generalization \cite{three}.  Physically, the vacuum's charged lepton pairs
which continuously fluctuate into and out of existence are acted upon by the
continuous distribution of virtual photons comprising the external electric
field; and if and when sufficient amounts of energy $E$ and momentum $p$ have
been transferred to each member of a pair, they become real leptons, each with
$E>p$.  Suppose now that charged, tachyonic degrees of freedom exist, perhaps
the same ones whose `` condensed '' summation defines the spontaneously broken
vacuum.  Then it is entirely reasonable to imagine that extremely strong
electric fields could transfer sufficient energy and momentum to the tachyon
pair, with each member here displaying $p>E$.  There is no conservation law that
forbids this; in principle, it is possible.

Of course, one immediate objection to $v>c$ motion is a resulting lack of
micro--causality.  But if the tachyon mass $M$ which sets the scale for acausal
effects is sufficiently large, e.g.  $M\sim 10^7$ GeV, this difficulty will not
be particularly relevant at typical laboratory distances.  Nevertheless, the
actions of charged tachyons after they are produced may well be unusual,
especially at galactic distances.

Mention should be made of the Wigner classification \cite{four} of relativistic
particles, wherein it is well known that a scalar tachyon field need have but
one component, but that the Little Group of `` higher dimensional '' tachyons is
given by the non trivial representations of $SU(1,1)$ which has only one
hermitian generator leading to a unitary representation in finite dimension,
instead of the three for usual massive particles for which the spin can be
defined and measured at rest.  In other words, for a spin $1/2$ massive
particle, the generators of the Little Group $SU(2)$ are $\{ \sigma_x, \sigma_y,
\sigma_z \}$, and the spin of the particle can be measured in any direction.  On
the other hand, for a two component tachyon ( for which no spin exists ), the
generators are $\{ i\sigma_x, i\sigma_y, \sigma_z \}$, and only one direction --
$z$ -- leads to a measurable quantity.  This loss of rotational invariance will
be manifest in the next paragraph.  Nevertheless, our model field theory
represents its leptonic tachyons in terms of four components, although somewhat
modified from those of Dirac, and the `` spin '' will be discussed in Appendix B.
  We also insist on the restriction to
orthochronous Lorentz transformations (LTs), so that -- in contrast to the work
of ref\cite{five} -- we absolutely forbid LTs which can interchange positive and
negative energy tachyons.  An amount of causality sufficient to define
scattering states requires a special assumption, but can be arranged.
Invariance under the full $CPT$ operation is maintained, although $P$ and $T$
invariance are separately violated.  And we stress that the micro-causality of
this model field theory is broken at very small space--time intervals.  But if
massive, charged tachyons provide a partial explanation of how the large scale
universe works, so be it.

Before writing any equations, we remind the reader of the Minkowski metric that
we are using :  $a_{\mu} = ( \vec a, ia_0 )$, so that $a^2 = \vec a^2 - a_0^2$.
Since the equations of General Relativity do not appear in this paper, there is
no distinction between covariant and contravariant indicies.  Our gamma matrix
notation is defined immediately after eq.  (2.1).

Several decades ago, other authors \cite{five} have discussed the possibility of
zero charge tachyons, and produced somewhat simpler models than the leptonic
tachyons suggested here.  Following the spirit of ref\cite{five}, it would be
possible to describe spin zero charged tachyons, which couple in a standard,
gauge invariant manner to the photon field, with an interaction term of the form
:  \be ieA_{\mu}\bigl(\varphi^{\dagger}\partial_{\mu}\varphi -
\partial_{\mu}\varphi^{\dagger}\varphi ) + e^2\varphi^{\dagger}A^2\varphi\ee
where $A_{\mu}$ denotes a conventional photon field operator.  What makes this
theory unsuitable for our astrophysical purposes is the
$e^2\varphi^{\dagger}A^2\varphi$ term of this interaction, which can describe
the simultaneous emission of a pair of photons; in this case the Cerenkov--like
angular restrictions of single photon emission and absorption are absent, which
restrictions are present when but a single photon is emitted by a tachyon.
Crucial to the astrophysical usefulness of the present model tachyons is the
strong probability that a very high energy tachyon can reabsorb every photon
that it emits, and thus retains its high energy across astrophysical times and
distances.  But if two photons were emitted simultaneously, the probability of
reabsorption of the photon moving in a direction roughly opposite to that of the
tachyon would be small.

The observant reader will shortly notice that our subsequent remarks are
 relevant to the interactions of such misgenerated tachyons with an
 electromagnetic field specified by a vector potential, which can represent
 either an external, classical field $A_{\mu}^{ext}(x)$, or the photon operator
 $A_{\mu}(x)$.  No reference will be made to a possible non linear coupling,
 e.g., ${\cal L}''=-(\lambda^2/2)(\bar\psi_T\gamma_5\psi_T)^2$, which would be
 appropriate in a Higgs or W-- S symmetry breaking context ( in which one
 neglects the gradient portions of the Lagrangian, as well as any
 electromagnetic coupling ).  But there is one electromagnetic situation in
 which such a ${\cal L}''$ might not be neglected, for a way must be found to
 block the decay of a mass shell photon into a pair of mass shell tachyons.  For
 ordinary, mass shell particles, the reaction $k\rightarrow p + p$ is
 kinematically impossible, whereas $k\rightarrow T + \bar T$ is, kinematically,
 perfectly possible.  Were this reaction not forbidden, perhaps by a selection
 rule, or severely damped, or given an energy threshold below which it is
 forbidden, any real photon of any energy would be able to decay into a $T\bar
 T$ pair, and our world would be quite different.

From a dynamical point of view, there are two possible sources of the needed
damping, or threshold effect.  One is the realization that the vertex
$k\rightarrow T + \bar T$ will resemble tachyon scattering by a photon, except
for the huge `` momentum transfer '' needed to replace one $T$ leg by a $\bar T$
leg; and, as such, serious, soft photon damping of this vertex may be expected.
The second source is damping induced by ${\cal L}''$, which, at least in eikonal
approximation, can be extracted from the radiative corrections that it
generates.

To make this argument more explicit, consider the set of Feynman graphs entering
into the amplitude $k\rightarrow T + \bar T$, which include the interactions
generated by ${\cal L}''$ along, each $T$ and $\bar T$ leg, and, in particular,
between $T$ and $\bar T$.  In effect, ${\cal L}''$ represents a $\lambda$
strength attraction between $T$ and $\bar T$, which must be overcome if the $T$
and $\bar T$ are to `` unbind ''.  The statement that it is difficult to break
them apart is, quantum mechanically, the statement that unless $k_0$ is
extremely large, the amplitude for this process is very small; and just this
effect is suggested by the ( non perturbative ) eikonal approximation mentioned
above.  Even more striking is a simple, cluster expansion approximation of such
radiative corrections to the Schwinger mechanism, which suggests that the
electric field needed for $T + \bar T$ production has an effective, $\lambda$
dependent threshold.

Calculations of these radiative corrections are underway, and will be presented
when completed.  In the meantime, we'll show - in section 5-A - using a simple
calculation, how the production of a $T \bar T$ pair by a real photon can be
forbidden, due to the presence of ${\cal L}''$.  In this paper, we shall only be
concerned with the electromagnetic interactions of such tachyhons after they are
produced.

The arrangements of these remarks is as follows.  In the next Section, a ``
misgenerated lepton '' model of tachyons is defined, and its spinorial content
outlined.  In Section 3, a complete, functional Quantum Tachyon Dynamics ( QTD )
is set up, with coupling to the electromagnetic field, in analogy to QED.  In
the next Section, the possibility of photon emission and reabsorption by charged
tachyons is discussed, and in Section 5 the kinematics of tachyon particle
reactions is outlined.  A version of Ehrenfest's theorem is exhibited in Section
6, to set the stage for a subsequent `` Classical Tachyon Dynamics '',
describing the classical motion of a charged tachyon in an external
electromagnetic field.  A final Summary and two Appendices complete the
presentation.  
\bigskip {\bf\section{QTD as a QFT}} \setcounter{equation}{0} 
The QFT here called QTD is modeled as closely as possible upon QED, with but two
changes, and one change of interpretation.  The standard QED Lagrangian may be
written as :  \be{\cal L}\,=-\,\bar\psi\,\Bigl[\, m + \gamma\!\cdot\!(\partial -
ieA)\,\Bigr]\psi + {\cal L}_0(A)\ee where ${\cal L}_0(A)$ denotes the free
photon Lagrangian, and $m$ and $e$ are the electron bare mass and charge,
respectively; in all of the following, hermitian $\gamma_{\mu}$ satisfying
$\{\gamma_{\mu}, \gamma_{\nu}\}= 2\delta_{\mu\nu}$ will be used, with
$\{\gamma_{5}, \gamma_{\mu}\}= 0$, $\gamma_j=-i\gamma_4\alpha_j$,
$\alpha_j=\sigma_1\otimes\sigma_j$, $\gamma_5=\gamma_1\gamma_2\gamma_3\gamma_4$,
where the elements of $\gamma_{4,5}$ are two--by--two matrices :
$\gamma_4=\pmatrix{1&0\cr0&-1\cr}$, $\gamma_5=-\pmatrix{0&1\cr1&0\cr}$.  The
$\sigma_j$ denote the hermitian Pauli matrices, satisfying
$\sigma_i\sigma_j=\delta_{ij}+i\epsilon_{ijk}\sigma_k$.  We use the Minkowski
metric~:  $x_{\mu}=(\vec x, ict)$, with $\hbar=c=1$ and
$\bar\psi=\psi^{\dagger}\gamma_4$.

From (2.1) and the Schwinger Action Principle, one obtains both the field
equations :  \be\Bigl[\, m + \gamma\!\cdot\!(\partial - ieA)\,\Bigr]\psi\,=0\ee
and, from the surface--term variations of the Action integral of (2.1), the
equal time anti commutation relation ( ETAR ) :  \be \Bigl\{\psi_{\alpha}(x),
\bar\psi_{\beta}(y)\Bigr\}_{x_0=y_0}\!\!\!\!\!\!\!=\gamma_4^{\alpha\beta}\,\delta(\vec
x - \vec y)\ee with (2.3) providing a statement of micro--causality, as expected
in a causal theory.

The first, and simplest, departure from QED that shall be adopted here is the
replacement of $m$ by $iM$, where $M$ is understood to be large, $M\gg m$~:
\be{\cal L} = -\,\bar\psi_{T}\,\Bigl[\, iM + \gamma\!\cdot\!(\partial -
ieA)\,\Bigr]\psi_T + {\cal L}_0(A)\ee This is not the final tachyon Lagrangian,
but only the simplest, initial form written to display the needed
$m^2\rightarrow -M^2$ continuation.  For simplicity, we allow the tachyon field
to be coupled to the electromagnetic field with the QED coupling constant, and
first consider the properties of the free tachyonic field, $\psi_T^{(0)}$,
satisfying :  \be(\,iM +
\gamma\!\cdot\!\partial\,)\,\psi_T^{(0)}\,=0\,=\bar\psi_T^{(0)}(\,iM +
\gamma\!\cdot\!\overleftarrow\partial\,)\ee One immediately sees that it will
not be possible to build a conserved current from $\psi_T^{(0)}$ and
$\bar\psi_T^{(0)}$ obeying equation (2.5).

In direct imitation of the free lepton fields' representations in term of
creation and destruction operators, and spinorial wave functions, we write :
\be\displaystyle\psi_T^{(0)}(x)=(2\pi)^{-3/2}\sum_{s=1}^2\int
\!\!d^3\!p\left[{iM\over E(p)}\right]^{1/2}\,\left\{b_s(\vec
p)\,e\,^{\displaystyle ip\!\cdot\!x}u_s(p) + d_s^{\dagger}(\vec
p)\,e\,^{\displaystyle -ip\!\cdot\!x}v_s(p)\right\}\ee where $b_s(\vec p)$ and
$d_s^{\dagger}(\vec p)$ are the destruction and creation operators,
respectively, for $T$ and $\bar T$, while $u_s(p)$ and $v_s(p)$ are their
appropriate spinors; and $p\!\cdot\!x=\vec p\!\cdot\!\vec x - E(p)x_0$, with
$E(p)=\sqrt{\vec p^2 - M^2}$.  Perhaps the simplest description appears if
$b_s(p)$ and $d_s(p)$ satisfy exactly the same anticommutation rules as for
ordinary leptons.  One then realizes that the $\exp\left(\pm ix_0\sqrt{\vec p^2
- M^2}\right)$ factors of (2.6) will generate exponentially growing or damping
time dependence of the $\psi_T^{(0)}$, $\bar\psi_T^{(0)}$, which is simply not
acceptable.  The obvious way to prevent this is to restrict $\vert\vec p\vert$
values to be larger than $M$, a solution which will have other consequences, but
ones that are acceptable and reasonable in a non causal theory.  Henceforth, a
factor of $\theta(\vert\vec p\vert-M)$ will appear under the momentum integrals
of $\psi_T^{(0)}$ and $\bar\psi_T^{(0)}$.

We shall give in Appendix B ( formula B.5) another, equivalent, form for 
the wave function.

Consider now the momentum--space equations which the tachyon $T$ and
anti--tachyon $\bar T$ must satisfy :  \be(\,M +
\gamma\!\cdot\!p\,)\,u_s(p)\,=0\,,\ \ \ \ \ \ (\,M -
\gamma\!\cdot\!p\,)\,v_s(p)\,=0 \ee Solutions of (2.7) may be expressed in terms
of the projection operators
$\Lambda_{\pm}(p)=\displaystyle{M\pm\gamma\!\cdot\!p\over2M}$~:  \be u_s(p)\,=
\displaystyle{\sqrt{2M\over\vert\vec p\vert}}\,\Lambda_-(p)\,\xi_s\,,\ \ \ \ \ \
v_s(p)\,=\displaystyle{\sqrt{2M\over\vert\vec p\vert}}\, \Lambda_+(p)\,\xi_{s+2}
\ee where, as for leptons, $s=1$ or 2, and $\displaystyle{\sqrt{2M\over\vert\vec
p\vert}} $ is an appropriate normalization constant.  The adjoint spinors $\bar
u=u^{\dagger}\gamma_4$, $\bar v=v^{\dagger}\gamma_4$, are then given by :  \be
\bar u_s(p)\,=\displaystyle{\sqrt{2M\over\vert\vec p\vert}}\,
\bar\xi_s\,\Lambda_+(p)\,,\ \ \ \ \ \ \bar v_s(p)\,=\displaystyle{\sqrt{2M\over\vert\vec p\vert}}\, \bar\xi_{s+2}\,\Lambda_-(p)
\ee and one sees that the norms of these states have the unfortunate property of
vanishing :  $\bar u_s(p)\,u_s(p)=\bar v_s(p)\,v_s(p)=0$.

Does this signify that it is impossible to construct lepton--style tachyonic
states ?  Not necessarily, because, -- as Dirac pointed out, long ago
\cite{fiveb} -- one can define expectation values with the aid of an indefinite
metric operator, $\eta=\eta^{\dagger}$, such that the proper norm of a state is
given by $\bar u\eta u$, rather than $\bar uu$.  In our case, there exists a
clear candidate for this metric operator :  $\eta=\gamma_5$, which has the
pleasant property of converting $\Lambda_{\pm}$ to $\Lambda_{\mp}$ as it passes
through either operator.  However, there is a price to pay for this convenience,
in that such norms are no longer scalars but pseudoscalars; and the vectors
constructed from them will be pseudovectors, etc.  It is difficult to avoid the suspicion
that such tachyons may turn out to be associated with the electroweak
interactions, forming the `` condensate '' of the W-- S vacuum, although that
subject will not be treated in this paper.

Following the above discussion, we shall insist that a factor of $\gamma_5$ be
inserted immediately after each $\bar u_s$ or $\bar v_s$ factor, for all matrix
elements subsequently calculated.  With this intention, the tachyonic part of
the Lagrangian of (2.4) shall now be written as :  \be{\cal
L}_T\,=-\,\bar\psi_{T}\,\gamma_5\,\Bigl[\, iM + \gamma\!\cdot\!(\partial -
ieA)\,\Bigr]\psi_T \ee with field equations :  \be\Bigl[\, iM +
\gamma\!\cdot\!(\partial - ieA)\,\Bigr]\psi_T
\,=0\,=\bar\psi_T\gamma_5\,\Bigl[\, iM - \gamma\!\cdot\!(\overleftarrow\partial
+ ieA)\,\Bigr]\ee so that the $\gamma_5$ insertion is automatic.  Were the
spinors $\xi_s$ chosen as for the leptons, $\xi_s=\chi_s$, where $\chi_1$ is the
column matrix of elements $(1, 0, 0, 0)$, $\chi_2$ has elements $(0, 1, 0, 0)$,
etc, then one easily calculates :  \be \bar
u_{s'}(p)\gamma_5u_s(p)\,=i\chi_{s'}^{\dagger}\vec\sigma\!\cdot\!\hat p\,\chi_s\
= i(\vec\sigma\!\cdot\!\hat p)_{s's}\ \ \ {\rm with}\ \ \ \ \hat p=\vec
p/\vert\vec p\vert\ee which, while non zero, does not display the usual
$\delta_{ss'}$, but corresponds to a linear combination of the spin states
labeled by $s$ and $s'$.

Similary, one has :  $ \bar
v_{s'}(p)\gamma_5v_s(p)\,=-\,i(\vec\sigma\!\cdot\!\hat p)_{s's}$.

The closure, or completeness relation for such spinors is then given by :
\be\sum_{s=1}^2\left\{u_s^{\alpha}(p)(\bar u_{s'}(p)\gamma_5)^{\beta} -
v_s^{\alpha}(p)(\bar v_{s'}(p)\gamma_5)^{\beta}\right\}(\vec\sigma\!\cdot\!\hat
p)_{s's}\,=\delta^{\alpha\beta} \ee The free field spinors now satisfy :  $$(\,M
+ \gamma\!\cdot\!p\,)\,u_s(p)\,= (\,M - \gamma\!\cdot\!p\,)\,v_s(p)\,=0$$ and :
$$\bar u_s(p)\gamma_5(\,M + \gamma\!\cdot\!p\,)\,=\bar v_s(p)\gamma_5(\,M -
\gamma\!\cdot\!p\,)\,=0$$ It should be noted that the Action formed from the
Lagrangian of (2.10) is hermitian, as is the Hamiltonian.  These assertions
require the customary neglect of spatial surface terms, as well as the
observation -- obtain from the field equations -- that :
$\displaystyle{\partial\over\partial t}\int\!\!
d^3x\,\bar\psi_T\,\gamma_5\gamma_4\psi_T\,=0$.

If the coefficient of $A_{\mu}$ in (2.10) is defined as the charged tachyon
current operator :  $J_{\mu}^T = ie\bar\psi_T\,\gamma_5\gamma_{\mu}\psi_T$, then
it follows from the field equations (2.11) that this charged current is
conserved :  $\partial_{\mu}J_{\mu}^T=0$.  Charged conjugation may be effected
by the same set of Dirac matrices used for the lepton case :
$\psi_c=C\bar\psi^T$, where the superscript $T$ denotes transposition and
$C=i\gamma_2\gamma_4$.

Finally, one must discuss the equal time anticommutation relation $
\Bigl\{\psi_T^{\alpha}(x), \bar\psi_T^{\beta}(y)\Bigr\}_{x_0=y_0}$, which -- as
for ordinary leptons -- are assumed to be the same for free and interacting
field operators.  The Schwinger Action Principle \cite{six} underlying QFT
specifies the ETAR by identifying the infinitesimal generators from the surface
terms of the variation of the Action~:  $\delta W=\displaystyle-\!\int\!\!
d\sigma_{\mu}\bar\psi\gamma_{\mu}\delta\psi$ in QED, or :  $\delta
W=\displaystyle-\!\int\!\!
d\sigma_{\mu}\bar\psi_T\gamma_5\gamma_{\mu}\delta\psi_T$ for the present case of
QTD.  For the usual case where the relevant space--time surface is taken as a
flat time cut, one has in QED :  $$\delta W=\displaystyle-\!\int\!\!
d\sigma_{4}\bar\psi\gamma_{\mu}\delta\psi=i\!\int\!\!  d^3y\,\bar\psi(\vec y,
t)\gamma_{\mu}\delta\psi(\vec y, t)$$ thereby identifying the field momentum
operator conjugate to $\psi_{\alpha}$ as $\pi_{\alpha}(\vec y,
t)=\left(\bar\psi(\vec y, t)\gamma_4\right)_{\alpha}$ and from this follows the
choice of ETAR :  $\Bigl\{\psi_{\alpha}(x),
\psi_{\beta}^{\dagger}(y)\Bigr\}_{x_0=y_0}\!\!\!\!\!\!\!=\delta_{\alpha\beta}\,\delta(\vec
x - \vec y)$, which expresses the fermion micro--causality of conventional QED.

For an acausal theory, one should not expect to require strict micro--causality;
but, keeping to the original statement of the Action Principle, it is possible
to obtain a modified expression of the QTD ETAR, which can be written in the
form :  \be\Bigl\{\psi_T^{\alpha}(x),
\left(\bar\psi_T(y)\gamma_5\right)^{\beta}\Bigr\}_{x_0=y_0}\!\!\!\!\!\!\!=(\gamma_4)^{\alpha\beta}\,\hat\delta(\vec
x - \vec y)\ee It is then straightforward to employ (2.13) and show that the
substitution of (2.6) and its adjoint into the LHS of (2.14) yields :
\be\hat\delta(\vec x - \vec y)=\int\!\!{d^3p\over(2\pi)^3}\,\theta(\vert\vec
p\vert - M)\,e\,^{\displaystyle i\vec p\!\cdot\!(\vec x - \vec y)}\ee which is
just the form one expects with the restriction $\vert\vec p\vert>M $.  This ``
modified delta function~'' of (2.15) is a well defined distribution of
$\vert\vec x - \vec y\vert$, which can be written in various form, such as :
\be\hat\delta(\vec x - \vec y)=\delta(\vec x - \vec y) -
\int\!\!{d^3p\over(2\pi)^3}\,\theta(M - \vert\vec p\vert)\,e\,^{\displaystyle
i\vec p\!\cdot\!(\vec x - \vec y)}\ee where the second RHS of (2.16) is a finite
measure of the lack of micro--causality of this tachyon theory.  With $\vec
z=\vert\vec x - \vec y\vert$, it is immediately evaluated as :  \be{M\over
2\pi^2z^2}\left\{{\sin(Mz)\over Mz} - \cos(Mz)\right\}\ee For small and large
$Mz$, (2.17) reduces to $\displaystyle{1\over6\pi^2}M^3$, $Mz\ll 1$; and to
$-\displaystyle{M\over 2\pi^2z^2}\cos(Mz)$, $Mz\gg 1$.  In other words, the lack
of micro--causality occurs mainly for extremely small distances, $z<M^{-1}$,
where it is significant, while, for $z>M^{-1}$, the effect is oscillatory and of
unimportant magnitude.  The basic theory, however, is non causal in a
qualitatively different sense, with tachyon propagation emphasized outside the
light cone but damped away inside, in exactly the opposite sense to that of QED.

The free tachyon propagator, defined as :  \beq{}&\displaystyle
S_T^{\alpha\beta}(x-y)=i<\left(\psi_T^{\alpha}(x)\left(\bar\psi_T(y)\gamma_5\right)^{\beta}\right)_+>\nonumber\\&\equiv\displaystyle
i\theta(x_0-y_0)\!<\psi_T^{\alpha}(x)\left(\bar\psi_T(y)\gamma_5\right)^{\beta}\!>-\,i\theta(y_0-x_0)\!<\left(\bar\psi_T(y)\gamma_5\right)^{\beta}\psi_T^{\alpha}(x)\!>\eeq
is then, with (2.14), to satisfy :  \beq{}&\displaystyle
(iM+\gamma\!\cdot\!\partial)\,S_T(x-y)=\delta(x_0-y_0)<\left\{\left(\gamma_4\psi_T(x)\right)^{\alpha},
\left(\bar\psi_T(y)\gamma_5\right)^{\beta}\right\}>\nonumber\\&\displaystyle
=\delta(x_0-y_0)\,\hat\delta(x-y)\nonumber\eeq and has the integral
representation :  \be S_T(z)=-i\!\!\int\!\!{d^4p\over(4\pi)^4}\,\theta(\vert\vec
p\vert - M){e\,^{\displaystyle ip\!\cdot\!z}\over(M+\gamma\!\cdot\!p\,)}\ee An
immediate question is the method of choosing contours to perform the momentum
integrals of (2.19), since :
$(M+\gamma\!\cdot\!p\,)^{-1}=(M-\gamma\!\cdot\!p\,)/(M^2-p^2)^{-1}$.  If, in the
usual way, $M^2\rightarrow M^2\pm i\epsilon$, one must decide which possibility
is appropriate.  Calculation of the $\displaystyle\int\!\!dp_0$ yields :  \be
S_T(z)=(\mp){1\over2}\int\!\!{d^3p\over(2\pi)^3}\,{\theta(\vert\vec p\vert -
M)\over\sqrt{\vec p^2-M^2}}\,\,e\,^{\displaystyle i\vec p\!\cdot\!\vec z\mp
i\vert z_0\vert\sqrt{\vec p^2-M^2}}\!\left( M - \vec\gamma\!\cdot\!\vec p\pm
i\gamma_4\epsilon(z_0)\right)\ee and if the $z\rightarrow0$ limit of (2.20) is
compared with a direct computation of the second line of (2.18), using the
completeness statement (2.13), the two calculations will agree if the upper sign
is used in (2.20); that is :  $M^2\rightarrow M^2+i\epsilon$.  For $z\ne0$, one
then finds agreement with the form of the fermion propagator in QED, in the
sense that an exponential factor $\exp( i\vec p\cdot\vec z-i\omega\,\vert
z_0\vert)$ is present in both theories, with $\omega=\sqrt{\vec p^2+m^2}$ for
QED and $\omega=\sqrt{\vec p^2-M^2}$ for QTD.  Henceforth, $M^2\rightarrow
M^2+i\epsilon$.

Because of the $\theta(\vert\vec p\vert - M)$ factor, one might guess that the
propagator of (2.19) and (2.20) is not Lorentz invariant (LI), but this turns
out not to be the case; in fact, the propagator is LI, and is related to the
conventional particle propagator by the simple continuation :  $m\rightarrow
iM$.  The easy way to see this is to set $S_T(z)=(iM-
\gamma\!\cdot\!\partial)\,\Delta_T(z, M)$, in analogy to the ordinary lepton
case, set $S_c(z)=(m- \gamma\!\cdot\!\partial)\,\Delta_c(z, m)$, and to insert a
factor of unity÷:  $1 = \theta(\vec p^2 + m^2)$, under the integrals that define
$\Delta_c(z, m)$.  If the continuation $m\rightarrow iM$ is now made, one finds
that :  \be\Delta_c(z, iM)=\Delta_T(z, M )
=-\int\!\!{d^4p\over(2\pi)^4}\,\theta(\vert\vec p\vert - M){e\,^{\displaystyle
ip\!\cdot\!z}\over p^2-M^2-i\epsilon}\ee a completely LI result in which the
space-like and time-like regions of $\Delta_c(z, m)$ have been interchanged.
Explicit calculation shows that the integral representation of $(2.21)$ does
indeed generate precisely $\Delta_c(z, iM)$.  
\bigskip {\bf\section{Functional QTD}} \setcounter{equation}{0} 
In this section, a generating functional for QTD
is constructed, in analogy with that of ordinary QFT, with charged tachyons
replacing charged leptons.  Because the properties of the free tachyon creation
and destruction operators have been chosen to be the same as those of ordinary
lepton fields, the transition from $n$--point functions of QTD to corresponding
$S$--Matrix elements can be taken over directly from ordinary QED.  However,
because of the ubiquitous $\theta(\vert\vec p\vert - M)$ factors, one requires
the understanding that tachyons $IN$ and $OUT$ field operators must be defined
at `` asymptotic times '' that are outside the light cone of any origin of
coordinates in terms of which the field operators might be evaluated.

For the Lagrangian :  \be {\cal L}\,=-\,\bar\psi_T\gamma_5\,\Bigl[ iM +
\gamma\!\cdot\!\left(\partial - ie[A+A^{ext}]\right)\,\Bigr]\psi _T+ {\cal
L}_0(A)\ee where $A$ denotes the ordinary, quantized, photon field, and ${\cal
L}_0(A)$ is its free Lagrangian (~in an appropriate gauge ), one defines the
generating functional in terms of tachyonic and photon sources as :  \be
<S[A^{ext}]\!>{\cal Z}\left[\eta, \bar\eta,
j\,\right]=<S[A^{ext}]\bigg(e\,^{\displaystyle
i\int\left[\bar\eta\psi_T+\bar\psi_T\eta+j_{\mu}A_{\mu}\right]}\bigg)_+\!\!>\ee
just as one would do for QED.  With the aid of the field equations, the photon
ETCR and the tachyonic ETAR -- or, directly, from the Schwinger Action Principle
-- one obtains \cite{six} :  \beq{}&\displaystyle{\cal Z}\left[\eta, \bar\eta,
j\,\right]=e\,^{\displaystyle {i\over 2}\int
\!\!jD_cj}\!\cdot\,e\,^{\displaystyle -{i\over 2}\int\!\!  {\delta\over\delta
A}D_c{\delta\over\delta A}}\cdot\,e\,^{\displaystyle i\!\int\!\!\bar\eta\,
G_T[A+A^{ext}]\gamma_5\,\eta}\nonumber\\&\cdot\,\,e\,^{\displaystyle
L[A+A^{ext}]}/\!\!<S[A^{ext}]\!>\eeq where, for (3.3),
$A_{\mu}=\displaystyle\int \!\!d^4y D_{c,\mu\nu}(x-y)j_{\nu}(y)$.  Because of
the normalization requirement ${\cal Z}[0,0,0]=1$, there follows :
\be<S[A^{ext}]\!>\,\,=e\,^{\displaystyle -{i\over 2}\int\!\!  {\delta\over\delta
A}D_c{\delta\over\delta A}}\cdot\,e\,^{\displaystyle
L[A+A^{ext}]}\left\vert_{A\rightarrow 0}\right.\ee where :  \be
L[A]\,=i\int_0^e\!de'\,{\rm Tr}\left[\gamma\!\cdot\!A\,G_T[e'A]\right]\ee and :
\be G_T[eA]\,=S_T\left[ 1 - ie(\gamma\!\cdot\!A)S_T\right]^{-1}\ee Because the $\gamma_5$ factors always appear with
a corresponding $\bar\psi_T$ in (2.14) and (3.3), there is no $\gamma_5$
presence in the log of the tachyon determinant of (3.5), nor in the Green's
function of (3.6).  This means that the mixing of vector and axial vectors
interactions, alluded to in Section 2, will not occur in this relatively simple
version of QTD.  

One can here define `` retarded '' and `` advanced '' tachyonic Green's function
:

$S_{R,A}=(iM-\gamma\!\cdot\!\partial)\Delta_{R,A}$, with :
\be\Delta_R(z)\,=\int\!\!{d^4p\over(2\pi)^4}\,{\theta(\vert\vec p\vert -
M)\,{e\,^{\displaystyle ip\!\cdot\!z}}\over p^2-M^2-i\epsilon s(p_0)}\ee where
$s(p_0)=p_0/\vert p_0\vert$; a similar representation, but with the sign of
$s(p_0)$ reversed, defines $\Delta_A(z)$.  Both Green's functions satisfy :
$(M^2+\partial^2)\Delta_{R,A}(z)=\delta(z_0)\hat\delta(\vec z)$, so that the
$S_{R,A}$ satisfy :
$(iM+\gamma\!\cdot\!\partial)S_{R,A}(z)=\delta(z_0)\hat\delta(\vec z)$.  Both
$\Delta_R$ and $S_R$ vanish for $z_0<0$, and hence merit the subscript $R$; and
conversely for $\Delta_A$ and $S_A$, which vanish for $z_0>0$.  This formalism
permits physical significance to be maintained for orthochronous Lorentz
transformations, with a well defined sense of time and of a particle's sign of
energy, while restricting consideration to Lorentz transformations that are
performed outside the light cone.

From the above discussion, the initial step of Symanzik's derivation
\cite{seven} of the functional reduction formula, between every $S$--Matrix
element and the generating functional, is valid.  It is :
\be\psi_T(x)\,=\sqrt{Z_2}\,\psi_T^{IN}(x) + \int\!\!d^4y\,S_R(x-y)\,{\cal
D}_y\,\psi_T(y)\ee where ${\cal D}_y=iM+\gamma\!\cdot\!\partial_y$, $M$ is the
renormalized ( physical ) tachyon mass, $\psi_T(x)$ denotes the fully
interacting ( Heisenberg representation ) tachyon operator, and $\psi_T^{IN}$ is
its free field counterpart, bearing its renormalized mass.  $Z_2$ is the tachyon
wave function renormalization constant.  Since ${\cal D}_x\,\psi_T^{IN}=0$, the operation of ${\cal D}_x$ on both sides of (3.8) yields a
simple identity.  Once the validity of (3.8) is appreciated, all of Symanzik's
functional steps follow through, with the result :  \beq{}&\displaystyle
S/\!\!<S[A^{ext}]\!>\,=\,:\,\exp \left\{
Z_3^{-1/2}\!\!\int\!\!A_{\mu}^{IN}(-\partial^2){\delta\over \delta
j_{\mu}}\right.\nonumber\\&\displaystyle\left.  +\, Z_2^{-1/2}\!\!\int\!\!\left[
\bar\psi_T^{IN}\gamma_5\,\vec{\cal D}{\delta\over \delta\bar\eta} - {\delta\over
\delta \eta}\overleftarrow{\cal D}\gamma_5\,\psi_T^{IN}\right]\right\}:\,{\cal
Z}\left[\eta,\bar\eta,j\right]\left\vert_{\eta,\bar\eta,j\rightarrow0}\right.\eeq
With (3.9), it is possible, in principle, to calculate the amplitude of any
process to any perturbative order; in addition, and as in certain limiting
situations in ordinary QFT, it may be possible to sum subsets of classes of
Feynman graphs, with each class containing an infinite number of graphs.
\bigskip {\bf\section{Photon Emission and Reabsorption}}
\setcounter{equation}{0} There is one important difference between the Physics
of certain `` modified bremsstrahlung~'' processes in QED and QTD which is worth
mentioning.  In conventional $S$--Matrix descriptions of reactions initiated by
a few particles, the initial and final states are understood to be
asymptotically well separated, and this is expected in a world where all massive
particles travel with $v<c$.  But when charged tachyons and photons are
involved, the situation is not as clear, because physical `` overlaps '' of
these particles can exist for macroscopic times; and the final, physical result
can be quite unexpected.

As an example, consider the amplitude for a charged tachyon of momentum
$T_{\mu}$ to emit a photon, and to leave the scene of the reaction with momentum
$T_{\mu}'$.  Kinematically, this process resembles Cerenkov radiation, with
$T=T'+k$.  Squaring and summing both sides of the relation $T' = T - k$, with
$T^2=T'^2=M^2$, $k^2=0$, leads to the restrictions :  $T\!\!\cdot\!k=\vec
T\!\!\cdot\!\vec k-T_0k_0=0$.  In other words, the angle between the photon's
spatial momentum $\vec k$, and the direction, $\hat T$, of the emitting tachyon,
is given by $\cos\theta=T_0/\vert\vec T\vert$; and similarly,
$\cos\theta'=T_0'/\vert\vec T'\vert$ defines the angle between the outgoing
$\vec k$ and $\vec T'$.  Note that these kinematical relations are independant
of the energy of the emitted photon.  Using standard Rules, as sketched in
Appendix A, one can calculate the probability per unit time $\tau$ for the
emission of a photon of arbitrary polarization within a band of energy $\omega$
to $\omega + d\omega$, as :  \be {1\over \tau}\,{d\over
d\omega}\,\sum\left\vert<T'\!,k\,\vert S\,\vert \,T\!>\,\right\vert^2\,={\alpha
M^2\over \vert\vec T\vert T_0}\ee where $\alpha=e^2/4\pi$.  ( A differential
cross--section is not appropriate here, since the angular distributions are
fixed by the kinematics above ).  With a physical upper cut--off chosen as
$\omega_{max}=T_0$, the lowest order probability/time for such emission is :
\be{1\over \tau}\,\sum\left\vert<T'\!,k\,\vert S\,\vert
\,T\!>\,\right\vert^2\,={\alpha M^2\over \vert\vec T\vert}\ee By direct
calculation, (4.2) has exactly the same probability/time for the inverse
process, the absorption of a photon $k$ by a tachyon $T$ which results in $T'$.
In order for this inverse process to become significant, the incident tachyon
should move in a directed photon beam, at just the right kinematical angle, for
an amount of time $\tau=\vert\vec T\vert/\alpha M^2$, using the perturbative
result for the inverse process as indicative of that of a more precise
calculation.  Because the tachyon moves faster than $c$, the question to be
posed is whether the outgoing photon of the emission process stays `` in close
proximity '' to the outgoing tachyon for a duration of time of this magnitude,
and can therefore be reabsorbed by that outgoing tachyon.  The angles for
emission and absorption for the two processes are identical, and therefore the
reabsorption of that emitted photon is a kinematic possibility.  The question to
be answered is :  how long are the outgoing photon and tachyon `` in close
proximity '' ?

The relative velocity of emitted photon and final tachyon is the relevant
quantity.  With $c=1$ ( and $\hbar=1$ ), the velocity vector of the emitted
photon is $\hat k$, while that of the final tachyon is $\vec v_T'=dT_0'/d\vec
T'=\vec T'/T_0'$, with $T_0'=\sqrt{\vec T'^2-M^2}$.  The relative velocity of
the two is $\vec v_{rel}=\hat k-\vec T'/T_0'$.  Because of the spatial,
kinematical relations stated above, the projection of $\vec v_{rel}$ along the
photon's direction vanishes :  \be \hat k\!\cdot\vec v_{rel}\,=0\ee Also, the
projection of $\vec v_{rel}$ along the tachyon's direction is small :  \be\hat
T'\!\cdot\vec v_{rel}\,=\hat k\!\cdot\hat T' -{\vert \vec T'\vert\over
T_0'}\,={T_0'\over\vert\vec T'\vert} - {\vert \vec T'\vert\over
T_0'}\,=-{M^2\over T_0'\vert \vec T'\vert} \ee for a high energy tachyon,
$T_0'\gg M$ ( for which case $T_0\gg M$ as well ).  A better indication of their
relative velocity might simply be \be\vec v_{rel}^2\,=\left(\hat k - {\vec
T'\over T_0' }\right)^2\,=1 + {\vec T'^2\over T_0'^2} - 2{\hat k\!\cdot\vec
T'\over T_0'}\,=+{M^2\over T_0'^2}\ee again using the kinematic relation $\hat
k\!\cdot\vec T'=T_0'$.  During a time $\tau$ one would then expect the photon
and tachyon to `` move apart '' by a distance $D\sim \tau(M/T_0')$.  The
significance of eqs.  (4.3)--(4.5) is that for a high energy tachyon, with
$T_0\gg M$ and $T_0'\gg M$, the emitted photon and final tachyon separate very
slowly.  Using (4.5) as a measure of their velocity of relative separation, and
the time $\tau\sim \vert\vec T'\vert/\alpha M^2$ as a qualitative measure of the
time suggested by (4.2) for the probability of reabsorption to be $\sim O(1)$,
the distance of that separation grows to $D=v_{rel}\,\tau\sim{\vert\vec
T'\vert\over T_0'}/\alpha M\sim 10^{-18}$ cm, for $M\sim 10^7$ GeV and $T_0'\gg
M$.  If a value larger than 1/137 is used for $\alpha$, or a larger value for
$M$, this distance $D$ is even shorter.

Contrast this with the distance moved by the tachyon, of velocity $\vert\vec
T'\vert/T_0'$, during the same interval $\tau$ :  $D_{T'}=v_{T'}\,\tau\sim\vec
T'^2/\alpha M^2T'_0$, which is a factor $\vert\vec T'\vert/M$ larger than $D$.
The final, high energy tachyon therefore moves through a distance which is
considerably larger than the separation distance between it and the emitted
photon; and this suggests that the photon and final tachyon remain `` in close
proximity '', such that during this time the photon is reabsorbed by that
tachyon.  At least, there should be a non-zero probability of reabsorption; and
continued in this way, over asymptotic times and distances, there should be a
significant, non-zero probability that a very high energy tachyon will reabsorb
every photon that it emits, and thereby retain its original high energy.  If not
exactly, then almost so.

Can a charged tachyon moving through an environment of Cosmic Microwave Background
 photons absorb those
CMB photons which it meets at just the proper angle ?  There is no reason why
this cannot happen, and it reinforces the notion that a charged tachyon moving
through galactic space can achieve -- and will maintain -- an ever increasing
energy, with a velocity just slightly larger than $c$.

 Another argument leads to a similar conclusion.  Consider the external $\vec E$
and $\vec B$ fields produced by a charged tachyon moving at constant velocity.
Any external field can be represented as an integral over virtual photon modes,
with weighting depending on the nature of the particular external field.  If
that field vanishes when viewed by an observer at the Cerenkov--like angle, it
means that no photons are emitted at that angle.  But emission at that angle is
precisely the definition of a real photon, since then and only then can $k^2=0$;
if the $\vec E$ and $\vec B$ fields of a constant velocity tachyon do vanish
when viewed at this angle, such a classical result indicates that no real
photons can be emitted by a tachyon in a Cerenkov--like process.  A
straightforward calculation of those fields shows that, if assumed continuous,
they indeed must vanish, precisely and uniquely at that angle.\bigskip
{\bf\section{Kinematics of Tachyon--Particle Reactions}}
\setcounter{equation}{0} The conjecture of the preceeding paragraph underlies
the reactions of this Section, in which high energy tachyons are able to retain
their energy/momenta while traveling to galactic distances, where they
annihilate and/or scatter with photons and other, ordinary matter.

One does not have the freedom to make Lorentz transformations that `` simplify
'' calculations by moving to a Lorentz frame in which the $T_0$ component of one
tachyon vanishes, for such a transformation assigns that tachyon an infinite
velocity.  This does not make any physical sense; the mathematical equivalent of
that step is dividing by zero.

Imagine that a charged $T\!\!-\!\bar T$ pair is produced at some space--time
point, either by a Schwinger mechanism following from quantized QED vacuum
fluctuations at extremely small distances \cite{one}, or by some other mechanism
associated with the Big Bang, or subsequent, cataclysmic events.  These tachyons
immediately separate, each with $v>c$, moving away from the world of ordinary
particles into the space between stars, and into the outer reaches of every
galaxy.  But galactic magnetic fields exist, which must influence these charged
tachyons in the conventional way, by bending their trajectories into partially
`` circular '' paths, depending on the direction and magnitude of the magnetic
fields encountered.  ( If there are a sufficient number of such tachyons, they
themselves can be thought of as generating at least a portion of the galactic
magnetic fields; but that subject is reserved for a separate treatment ).  The
picture that emerges is of `` swarms '' of charged tachyons, moving not quite
randomly in curved orbits at galactic distances; and every so often, depending
on the density of such tachyons and of stray photons and bits of ordinary
matter, there will occur collisions.  
\vskip0.3truecm 
A) Blocking the reaction
$\gamma\rightarrow T\!\!+\!\bar T$ :

\hskip1truecm We here suggest a simple, intuitive mechanism for preventing the
decay of real, mass--shell photons into $T\!\!-\!\bar T$ pairs, a mechanism
which may be related to the conventional Higgs--Weinberg--Salam symmetry
breaking.  One imagines a $\psi^4$ type contribution to the basic interaction
lagrangian, ${\cal L''} = -(\lambda^2/2)\bigl(\bar\psi\gamma_5\psi\bigr)^2$, or
a less singular relative ( such as :  $i\lambda\bar\psi\gamma_5\psi\chi -
1/2\,\bigl[ \,\mu^2\chi^2 + (\partial\chi)^2\, \bigr]$ ) which provides a
strong, short--range, binding interaction between $T$ and $\bar T$, such that
the excitation of a vacuum state $T\bar T$ pair by the Schwinger mechanism
requires the addition of a sizable energy, $B$, to overcome this assumed
$T\!\!-\!\bar T$ binding.  Were a photon able to lift a $T\!\!-\!\bar T$ pair
from the vacuum, it also would be required to supply $B$ into the conservation
equations :$$k_0 = B+T+\bar T\ ,\ \ \ \ \vec k = \vec T + \vec{\bar T}$$For
these mass--shell particles, $k^2 = \vec k^2 - k_0^2 = 0$, $T^2 = \bar T^2 =
M^2$; and if the angle between the supposedly emitted $T$ and $\bar T$ is
$\theta$, it is straightforward to show that :
\be\cos^2\theta\,={\bigl[\,T_0\bar T_0-M^2+B^2/2+B(T_0+\bar T_0)+(T_0^2 + \bar
T_0^2)/2\,\bigr]^2\over [M^2+T_0^2][M^2+\bar T_0^2]}\ee But, for example, if
$B=2M$, a little algebra shows that $\cos^2\theta>1$, for arbitrary $T_0$, $\bar
T_0$; and hence this process is forbidden.

A corollary to this argument is the observation that an extra amount of energy
$B$ is released by the annihilation reaction :  $T + \bar T\rightarrow\gamma +
\gamma$, as mentioned in connection with ``~loop annihilation '' at the close of
Section 6.  If $B\sim 2M$, there would be an additional and large amount of
energy released in each annihilation, in addition to the intrinsic $T_0$ and
$\bar T_0$ energies of each of the annihilating tachyon pair.

\vskip0.3truecm B) Scattering :  $T+p=T'\!+p'$

\hskip1truecm Suppose that a particle ( e.g.  a proton floating about in outer
galactic space ) is struck by an energetic tachyon.  We idealize this situation
to a proton at rest struck by a tachyon of energy $T_0\gg m$, write the
conservation laws in the form $T+p-p'=T'$, square and sum both sides to find an
exact, linear equation ( assuming $p_0'>m$ ) :  \be p'_0\,={N\over
D}\,m\,={(m+T_0)^2+(M^2+T_0^2)\cos^2\theta\over(m+T_0)^2-(M^2+T_0^2)\cos^2\theta}\,m\ee
where $\theta$ is the angle between $\vec T$ and $\vec p\,'$.  Since $N>0$, one
expects that the largest value of $p_0'$ will result from the smallest value of
$D$, or the maximum possible value of $\cos^2\theta$.  From energy conservation,
$p_0+T_0-p_0'=T_0'>0$, which when applied to the present case of $p_0=m$,
requires :  \be\cos^2\theta<{T_0(m+T_0)^2\over (M^2+T_0^2)(2m+T_0)}\,\equiv
\cos^2\theta_{max}<1\ee Substitution of the $\cos^2\theta_{max}$ of (5.3) into
(5.2) then yields :  \be p_0'\,=m\,\left[{1+\displaystyle{1\over 1+2m/T_0}\over
1-\displaystyle{1\over 1+2m/T_0}}\right]\,\simeq T_0\left(1 - {m\over
T_0}+\cdots\right)\ee An effective `` billiard ball '' collision has occurred,
with the final proton taking on the energy of the incidental tachyon.

This example shows that an initial tachyon is able, by scattering, to transfer a
considerable amount of its energy to a particle initially at rest ( or moving at
a modest $v<c$ ), and in the process create high energy cosmic rays.  Radiation
from the scattering of charged particles by this mechanism should, in principle,
contribute $X$--ray and $\gamma$--ray backgrounds.

In addition, here is a mechanism which explains how occasional ultra high energy
cosmic rays, specifically protons, can have energies exceeding the GZK limit
\cite{nine}.  A stray, high energy charged tachyon, approaching the Earth,
strikes and transfers a large portion of its energy and momentum to a low energy
proton above the atmosphere, which proton promptly initiates the observed cosmic
ray shower.  The GZK limit does not apply to the tachyon, while the proton
obtains its extremely high energy just before initiating that shower.  
\bigskip
{\bf\section{From Ehrenfest's Theorem to Loop Annihilation}}
\setcounter{equation}{0} 
It has been suggested above that charged tachyons of
very high energy could reabsorb any photon emitted, thereby preserving their
four--momentum across galactic times and distances; and, as they move away from
and between galaxies, such tachyons could be influenced by galactic magnetic
fields ( of orders of magnitude from milli to micro Gauss ) and caused to move
in `` circular '' paths, at least while they are under the influence of a
coherent magnetic field.

A fundamental question arises as to just how a charged tachyon would be affected
by specified electromagnetic fields.  There are no `` classical '' tachyons in
our world, upon which experiments could be performed to yield Classical Tachyon
Dynamics ( CTD ), and so the analysis must proceed in the opposite sense :  from
the assumed QTD Lagrangian of (2.10) to a one--particle representation of that
Lagrangian, and thence to a subsequent approximation recognized as representing
the `` classical '' action of a specified electromagnetic field on a single,
charged tachyon.  These last steps may be realized by a derivation of the
corresponding Ehrenfest's theorem, and that is the subject of this Section.

Begin first with the free tachyon field operator of (2.6), and consider its
one--particle matrix element :  $$\phi(x)\,=<0\,\vert\,\psi_T(x)\,\vert\,\vec
p,s\!>$$ which satisfies the same equation as does $\psi_T(x)$, and has its
solution ( see (B.8) ):  $$\phi(x)\,= e\,^{\displaystyle ip\!\cdot\!x}\,u_s(p)$$ 
Because $u_s(p)$ satisfies :
$(M+\gamma\!\cdot\!p)u_s(p)\,=0$, upon multiplication by $i\gamma_4$ one obtains
:  $$( i\gamma_4M+\vec\alpha\!\cdot\!\vec p\,\, )\,u_s(p)\,=p_0\,u_s(p)$$ which
permits the identification of $H_T^{(0)}=iM\gamma_4+\vec\alpha\!\cdot\!\vec p$
as the one--free--tachyon Hamiltonian operator, with real eigenvalue $p_0$.
Note that this Hamiltonian is not hermitian, which reflects the fact that part
of its spectrum is imaginary ( for $|\vec p|\le M$ ) :$${\rm sp}\,H_T^{(0)} =
{\rm I\!R}\cup [ -iM, iM ]$$ Here, of course, the $\theta()$ factors insure that
only real eigenvalues appear.

In QED, the same analysis provides the one--free--electron Hamitonian,
$H^{(0)}=m\gamma_4+\vec\alpha\!\cdot\!\vec p$.

The gauge invariant extension of this one--tachyon Hamltonian to include the
effects of an external electromagnetic field is immediate :  $$H_T\rightarrow
iM\gamma_4+\vec\alpha\!\cdot\!( \vec p - e\vec A ) + eA_0$$ just as one does in
QED.

The time dependence of the one--tachyon wave function, assumed properly
normalized according as :  $\int \!d^3x\,\phi^{\dagger}\gamma_5\phi=1$ is given
by the relevant Schr\"odinger equation $i\hbar\,{\partial\phi/\partial
t}=~H_T\phi$, and its adjoint $-i\hbar\,{\partial\phi^{\dagger}/\partial
t}=\phi^{\dagger}H_T^{\dagger}$, while the expectation value of any tachyonic
operator $Q$ is defined by :
$$<Q>\,=\int\!\!d^3x\,\phi^{\dagger}\gamma_5\,Q\,\phi$$ As usual in this ``
Schr\"odinger representation '', the full time dependence is carried by the wave
functions, and the operators $Q$ are understood to be time independent.  But the
$c$--number $A_{\mu}^{ext}(x)$ are allowed to depend upon time, since unitary
transformations on $c$--numbers cannot effect their time dependence.  Hence, the
total time rate of change of $<Q>$ is given by :  $${d\!<Q>\over
dt}\,=\int\!\!d^3x\,\phi^{\dagger}\gamma_5\,\left({\partial Q\over\partial
t}+{1\over i\hbar}[\,Q,H_T]\right)\phi$$ where, since $\{
\,\gamma_5,\gamma_4\,\}=0=[ \,\gamma_5, \vec \alpha \,\,]$,
$H_T=\gamma_5H_T^{\dagger}\gamma_5$.  Thus, if $Q\rightarrow\vec r$, $${d\!<\vec
r>\over dt}\,=\int\!\!d^3x\,\phi^{\dagger}\gamma_5\,\vec\alpha\,\phi$$ showing,
as in QED, that $\vec\alpha$ is the operator whose expectation value corresponds
to velocity.

The average momentum of a charged tachyon in an external electromagnetic field
is calculated from the gauge invariant, `` canonical '' contribution :  $$<\vec
p>\,=\int\!\!d^3x\,\phi^{\dagger}\gamma_5\,( \vec p - e\vec A )\,\phi$$ and one
easily finds that :  \beq{}\displaystyle {d\!<\vec p>\over
dt}&=&\displaystyle-e\int\!\!d^3x\,\phi^{\dagger}\gamma_5\,{\partial\vec
A\over\partial t}\,\phi + {1\over
i\hbar}\int\!\!d^3x\,\phi^{\dagger}\gamma_5\,\left[(\vec p - e\vec A),
H_T\right]\,\phi\nonumber\\&=&\displaystyle
e\int\!\!d^3x\,\phi^{\dagger}\gamma_5\,\left[ -\vec\nabla A_0-{\partial\vec
A\over\partial t}+ \vec\alpha\times\vec B\right]\,\phi\\&=&\displaystyle
e\int\!\!d^3x\,\phi^{\dagger}\gamma_5\,\left[ \vec E + \vec\alpha\times\vec
B\right]\,\phi\nonumber\eeq Exactly the same relation, but without the
$\gamma_5$ factor, is obtained in QED.

If we now suppose that $\vec E$ and $\vec B$ are spatially constant over the
dimensions of the tachyon -- or, more precisely, over the wave packet that
describe the tachyon, as it moves with $v>c$ -- then (6.1) can be rewritten as :
$${d\!<\vec p>\over dt}\,=e\left[\, \vec E +<\vec v>\!\times\vec B\,\right]$$
which is just the conventional relation for a charged particle moving in
specified $\vec E$ and $\vec B$ fields, with $\vec p$ and $\vec v$ now replacing
$<\vec p>$ and $<\vec v>$.  If we wish to write the corresponding four vector as
:  $$p_{\mu}\,=M\,{d\!x_{\mu}\over d\tau}\,=Mu_{\mu}$$ where
$\displaystyle\sum_{\mu}u_{\mu}^2=+1$ ( rather than the $-1$ of the Minkovski
metric ) and :  $${dp_0\over d\vec p}\,={\vec p\over p_0}\,=\vec v$$ the
covariant equation for the motion of a charged particle is again valid, as in
QED :  $${d^2x_{\mu}\over d\tau^2}\,={e\over M}\,F_{\mu\nu}(x){d\!x_{\nu}\over
d\tau}$$ except that the tachyon mass $M$ replaces the electron's $m$, and
$dt/d\tau=\gamma=1/\sqrt{v^2-1}$.

With this demonstration that the motion of a charged tachyon follows essentially
the same classical equations as those of other charged particles, it is now
possible to understand how such charged tachyons can contribute to dark matter.
Consider the production of a burst of charged $T$s, and $\bar T$s, by some
catastrophic process.  They separate, presumably moving in roughly opposite
directions; and, concentrating on the $T$s, assume they possess a distribution
of velocities moving roughly in the same direction.  The lead $T$ of this group,
moving with the fastest velocity, generates a magnetic field which it outruns
and cannot feel; but the next $T$ behind it will be affected by that field, and
deflected so that it eventually lines up behind the first $T$.  And the same
process happens all way down the line, resulting in a `` line '', or effective
current of such charged $T$s.

Imagine that this current line is moving through galactic space -- the space
between galaxies -- where there exist magnetic fields of magnitude $B\sim
10^{-6}$ Gauss, which are coherent across distances $R\sim 10^4$ light years.
One naturally expects such $T$s to fall into orbits defined ( reinstating the
$c$ dependence ) by :  \be Mv^2/R\,=\gamma^{-1}evB/c\ee where
$\gamma^{-1}=[(v/c)^2-1]^{1/2}$, and $v>c$.  For high energy $T$s, $v\ge c$, and
so let us choose $(v/c)^2=1+10^{-6}$, $\gamma=10^3$, and insert the observed $B$
and $R$ values to obtain from (6.2) an estimate of what we might expect for $M$.
Amusingly, the result is $M\simeq 10^7$ GeV, which is just the order of
magnitude found in the original calculation of ref\cite{one}.

The point of this argument is that if those $B$ fields arise from currents in
the visible galactic matter, then the charged $T$s trapped in orbits by those
$B$s are `` rigidly '' connected to the matter that we can see; but the $T$s
themselves are invisible, they are `` dark ''.  (~Note that were any coherent
cyclotron radiation possible by such loops, its typical frequency would be far
too small to detect, $\omega=v/R\simeq 10^{-14}$ s$^{-1}$ ).  If there are
enough such charged $T$s, they could contribute to the dark, or missing matter
of current observation.  This argument can be easily generalized to include $T$s
of all energies.

Further, if there exist two or more such loops, of roughly the same size,
parallel to each other and separated horizontally by a distance on the order of
their radius, then their own magnetic fields would combine with that of the $B$
which hold them to create a `` magnetic bottle '' effect, trapping any available
number of ordinary charged particles, which would then also contribute to dark
matter.  Two such loops, each with their `` mega--macroscopic~'' magnetic
moments pointing along the magnetic field $\vec B$ would form a stable system :
their magnetic moments attract but their like electric charges repel each other,
leading to an equilibrium separation distance of order of their loop radius.

Finally, if one imagines that one loop consists of $\bar T$ circulating in the
opposite direction to that of the $T$ loop, their loop magnetic moments point in
the same direction along the line separating them, and the $T$ and $\bar T$
loops would be attracted towards each other -- both their magnetic moments and
unlike charges attract -- eventually leading to a cataclysmic `` loop
annihilation '', which might even be visible astronomically.  If the $T$ and
$\bar T$ loops encircle a large gas cloud, as they well might, then the ``
perfect fusion~'', loop--annihilation could emit sufficient radiation and newly
formed high velocity matter to irradiate and compress the gas, setting off
internal shock waves, and perhaps even a subsequent nuclear reaction.  The
corresponding $\gamma$--ray production, with $X$--ray and optical tails, and a
natural afterglow, could possibly match the well known, observed GRB data
\cite{nine}.
\bigskip {\bf\section{Summary}} 
\setcounter{equation}{0} 
In the belief that any model which can simultaneously relate astrophysical/experimental
results involving vacuum energy, dark matter, and the origin of ultra high
energy cosmic rays and gamma bursts is worth considering, we have here put forth
the idea of charged tachyons, as a possible, underlying mechanism.  Our charged
tachyons are fermionic, rather than scalar, to minimize the probability of
Cerenkov--like radiation by the tachyons over galactic distances. 

If the tachyons do not couple directly to pions, but only through higher order
photonic interactions, then the Greisen-Zatsepkin-Kuzmin cut-off ( GZK
reference, in ref\cite{nine} ) of high-energy cosmic rays is effectively
removed, compared to the case of energetic protons; and only after a collision
of a very high energy tachyon with a slow proton -- resulting in that proton's
absorption of a sizeable fraction of that tachyon's energy -- does the standard
GZK cut-off come into play.  But if that newly accelerated proton lies in our
upper atmosphere, and fairly close to Earth, it can effectively violate the GZK
limit.

Precisely how much of a contribution to Dark Matter could be made by such
tachyons depends on the knowledge of several points not yet clear :  the rate of
charged tachyon pair production by the Schwinger mechanism, and possible sources
of electromagnetic fields to induce such production; the average number of
tachyons per galactic loop, and the average number of such loops per galaxy.
Estimates of some of these questions are now underway; but at the moment, all
that one can reasonably say is that such charged tachyons provide a mechanism
which could contribute to the understanding of three current, astrophysical
problems.  We hope that the QTD analysis sketched above will generate further
efforts in this direction.

In addition to the topics mentioned above, it would be of considerable interest
to understand, in detail, the following items.

1) Possible radiation signatures from the scattering of tachyons on ordinary
particles, at galactic distances, especially if the tachyon charge is different
from the charge of the particle it strikes.

2) Induced emission and absorption of CMB photons.

3) The use of these charged tachyons as a possible condensation--mechanism for
the Weinberg--Salam vacuum.

4) The incorporation of such tachyons into a General Relativistic framework.

One final remark ( in response to constructive criticism of J.  Avan ).  One may
feel uncomfortable without a proper definition of the tachyon vacuum state, in
terms of conventional particle theory.  Let us ask the familiar question :  what
is the mass of a quark?  The well known answer is that it is physically
impossible to measure directly its mass, its intrinsic energy; and that one must
be content with estimates that can be inferred from measurements which display
results that depend on the type of measurement.  If quantum mechanics teaches us
anything, it is that no statements should be made -- or can be believed -- about
quantities that cannot be measured.  But we infer that a quark must have an
intinsic mass, and we assume that a quark vacuum state exists.

In an analogous way, one cannot directly measure any properties of our assumed
tachyons, but they can be used to explain, or to contribute to the understanding
of certain, puzzling, astrophysical observations.  Does our lack of ability for
direct measurements mean that there exists no tachyon vacuum ?  Not at all; it
merely reflects the fact that it is impossible for us to measure directly the
properties of an isolated, individual tachyon, just as for the quark situation,
but for different kinematical and dynamical reasons.  Again, all we can do is to
infer the properties of tachyons by matching their predictions against
astrophysical observations; and, again, we can only assume the existence of a
tachyon vacuum.  
\vskip1truecm {\bf Acknowledgments} 
\vskip0.3truecm 
The authors warmly thanks Prof.  Ian Dell'Antonio for several, most informative
conversations and Dr.  Fr\'ed\'eric Daigne for detailed informations on GRB
observations.  Discussions with J.  Avan and W.  Becker were most helpful.

\vskip1cm \def\theequation{A.\arabic{equation}} {\bf\section*{Appendix A}}
\setcounter{equation}{0} \appendix A brief sketch of the probability/time ``
intrinsic bremsstrahlung '' production, to lowest perturbative order, eq.
(4.1), proceeds as follows.

The $S$--matrix element of the lowest order, free field, photon and tachyon
operators :
\be<p',k\,|\,S\,|\,p\!>\,=ie\!\int\!\!d^4x\!<p',k\,|\,:\bar\psi_T(x)\gamma_5\gamma\!\cdot\!A(x)\psi_T(x):\,|\,p\!>\ee
may be calculated by inserting into (A.1) the conventional photon operator, and
the tachyon field operator (2.6) and its adjoint, which yields :
\be<p',k\,|\,S\,|\,p\!>\,={ie\over\sqrt{2\pi}}\,\sqrt{M^2\over
EE'}{1\over\sqrt{2\omega}}\left(\bar u_{s'}(p')\gamma_5\gamma\!\cdot\!\epsilon\,
u_s(p)\right)\delta^{(4)}(p-p'-k)\ee where $\epsilon_{\mu}$ is the photon
polarization, and we use $\vec p$ and $E(p)=\sqrt{\vec p^2-M^2}$ to represent
the tachyon coordinate.  Upon calculating $|\!\!<p',k\,|\,S\,|\,p\!>\!|^2$, the
square of $\delta^{(4)}(p-p'-k)$ is, as always, replaced by $(2\pi)
V\tau\,\delta^{(4)}(p-p'-k)$.  The relevant quantity to calculate is then not
$|\!\!<p',k\,|\,S\,|\,p\!>\!|^2$ summed over all $\vec p$ and $\vec k$ values
and polarizations ( and averaged over initial and summed over final spin indices
), but the probability for this process to occur per unit time :
\beq{}\displaystyle {1\over
\tau}\sum|\!\!<p',k\,|\,S\,|\,p\!>\!|^2\,&=&\displaystyle{M^2e^2\over(2\pi)^2E}\int\!\!{d^3k\over2\omega}\,{\delta(E(\vec
p)-E(\vec p-\vec k)-\omega)\over E(\vec
p)-\omega}\nonumber\\&\displaystyle&\times\sum_{s,s'}{1\over 2}\,\vert \,\bar
u_{s'}(p-k)\gamma_5\gamma\!\cdot\!\epsilon\, u_s(p)\,\vert^2\eeq The spin and
polarization sums produce exactly unity, and the continuous energy $\delta$
function of (A.3) may be replaced by $(E-\omega)/(\omega
p)\,\delta(\cos\theta-E/p)$ where $\theta$ is the angle -- fixed by the
conservation laws -- between $\hat p$ and $\hat k$.  The result is :  $${\alpha
M^2\over pE}\int_0^{\omega_{max}}\!\!d\omega$$ and we choose the maximum
possible value of $\omega_{max}$ as $E$, yielding the result quoted in Section
4.
\vskip0.5cm 
\def\theequation{B.\arabic{equation}} {\bf\section*{Appendix B}}
\setcounter{equation}{0} \appendix In this appendix, we will use the more familiar 
metric $(+,-,-,-)$
\\\\ 
\indent I. Free tachyon equation. \\\\ The $\gamma$ matrices satisfy :$$\left\{
\gamma^{\mu}, \gamma^{\nu}\, \right\} = 2\,g^{\mu\nu}$$ 
In the standard representation :
$$\gamma^0 = \pmatrix{I&0\cr 0&-I\cr},\ \ \ \ \vec\gamma = \pmatrix{0&\vec\sigma\cr -\vec\sigma&0\cr},\ \ \ \ 
\gamma^5 = i\gamma^{0}\gamma^{1}\gamma^{2}\gamma^{3} = \pmatrix{0&I\cr I&0\cr}$$In the free case, 
the tachyon equation is given by :
\be\left( i\gamma^{\mu}\partial_{\mu} - iM\right)\psi (x)
=0\ee
and :
$$\psi^{\dagger}(x)\gamma^5\gamma^0\left( i\gamma^{\mu}\overleftarrow\partial_{\mu} +
iM\right) =0$$
From these two equations, a conserved current can be built :$$j^{\mu}(x) =
\psi^{\dagger}\gamma^5\gamma^0\gamma^{\mu}\psi(x) = j^{\mu\dagger}$$It follows that $j^0 =
\psi^{\dagger}\gamma^5\psi$. 
\\ The tachyon equation can be obtained from the lagrangian density :$${\cal L}(x) = \psi^{\dagger}(x)\gamma^5\gamma^0\left(
i\gamma^{\mu}\overrightarrow\partial_{\mu} - iM\right)\psi (x) = {\cal
L}^{\dagger}(x)$$ \\ One gets for the conjugate momentum field $\pi$:  $$\pi(x) = {\partial{\cal L}(x)\over\partial\dot\psi(x)} =
i\psi^{\dagger}(x)\gamma^5$$The hamiltonian density is then :
$${\cal H}(x) = \psi^{\dagger}(x)\gamma^5\gamma^0\bigl[ -i\vec\gamma\!\cdot\!\vec\nabla +
iM\bigr]\psi(x) = 
i\psi^{\dagger}(x)\gamma^5\dot\psi(x) = {\cal H}^{\dagger}(x)$$  
\\
\indent II. Plane wave solutions.
\\\\ 1) Positive energy solutions.\\ We set $\psi(x) = u(p)\, e\,^{\displaystyle -i
p\!\cdot\!\!x}$.  Inserted into  (B.1), it gives :\be\left( \slash \!\!\!\!p - iM
\right) u(p) = 0\ee The spectrum of $\slash \!\!\!\!p
- iM$ being $(-2iM,-2iM,0,0)$ there will be two independent solutions 
$u^{(1)}(p)$ and $u^{(2)}(p)$ for (B.2).  \\ They can be obtained by using the relation :
$\left( \slash \!\!\!\!p - iM \right)\left( \slash \!\!\!\!p + iM \right) = p^2
+ M^2 = 0$.\\  One has :$$\slash \!\!\!\!p + iM = E\gamma^0 - \vec p\cdot\vec\gamma + iM =
\pmatrix{E+iM&0&-p_z&-p_x+ip_y\cr 0&E+iM&-p_x-ip_y&p_z\cr
p_z&p_x-ip_y&-E+iM&0\cr p_x+ip_y&-p_z&0&-E+iM\cr}$$\indent Applying it
 to the two vectors $u_1 = \pmatrix{1\cr 0\cr 0\cr 0\cr}$ and $u_2
= \pmatrix{0\cr 1\cr 0\cr 0\cr}$, one obtains:  $$u^{(1)}(p) =
{1\over\sqrt{2iM(E+iM)}} \pmatrix{E+iM\cr 0\cr p_z\cr p_x+ip_y\cr}\ \ {\rm and}\
\ u^{(2)}(p) = {1\over\sqrt{2iM(E+iM)}} \pmatrix{0\cr E+iM\cr p_x-ip_y\cr
-p_z\cr}$$\indent The $u_1$ and $u_2$ vectors, and the normalization of
$u^{(1)}(p)$ and $u^{(2)}(p)$, have been chosen in order to mimic 
the standard solutions of the Dirac equation  ( see, for instance, Bjorken and Drell \cite{ten}
or Itzykson and Zuber \cite{eleven} ), the electron mass $m$ being simply replaced
 here by $iM$, the mass of the tachyon.
\\\\ 2) Negative energy solutions.\\ We set $\psi(x) = v(p)\, e\,^{\displaystyle i
p\!\cdot\!\!x}$.  Substituted into  (B.1), it gives :\be\left( \slash \!\!\!\!p + iM
\right) v(p) = 0\ee One notices that :
$$\left( \slash \!\!\!\!p + iM \right) \gamma^5u(p) = -\gamma^5\left( \slash
\!\!\!\!p - iM \right) u(p) = 0$$One then chooses $v^{(1)}(p) = \gamma^5u^{(1)}(p)$,
$v^{(2)}(p) = \gamma^5u^{(2)}(p)$; and obtains :  $$v^{(1)}(p) =
{1\over\sqrt{2iM(E+iM)}} \pmatrix{p_z\cr p_x+ip_y\cr E+iM\cr 0\cr}\ \ {\rm and}\
\ v^{(2)}(p) = {1\over\sqrt{2iM(E+iM)}} \pmatrix{p_x-ip_y\cr -p_z\cr 0\cr
E+iM\cr}$$\indent As for $u^{(1)}(p)$ and $u^{(2)}(p)$, the vectors
$v^{(1)}(p)$ and $v^{(2)}(p)$ have been chosen in order to mimic the standard negative energy 
solutions of the Dirac equation.
\\\\
\indent III. Some useful formulae, valid for $\vert\vec p\vert\ge M$ :
\\\\ $u^{{\dagger}(\alpha)}(p)u^{(\beta)}(p) =
\displaystyle{\vert\vec p\vert\over M}$ \\
$u^{{\dagger}(\alpha)}(p)\gamma^0u^{(\beta)}(p) = 0$ \\
$u^{{\dagger}(\alpha)}(p)\gamma^5u^{(\beta)}(p) = \displaystyle{E\over
iM}\,i(\vec\sigma\!\cdot\!\hat p)_{\alpha\beta}$ \\
$u^{{\dagger}(\alpha)}(p)\gamma^5u^{(\beta)}(p)(-i\vec\sigma\!\cdot\!\hat
p)_{\beta\alpha'} = \displaystyle{E\over iM}\,\delta_{\alpha\alpha'}$ \\
$u^{{\dagger}(\alpha)}(p)\gamma^5\gamma^0u^{(\beta)}(p) =
i(\vec\sigma\!\cdot\!\hat p)_{\alpha\beta}$ \\
$u^{{\dagger}(\alpha)}(p)\gamma^5\gamma^0u^{(\beta)}(p)(-i\vec\sigma\!\cdot\!\hat
p)_{\beta\alpha'} = \delta_{\alpha\alpha'}$ \\
$u^{(\alpha)}(p)(-i\vec\sigma\!\cdot\!\hat p)_{\alpha\beta}\otimes
u^{{\dagger}(\beta)}(p)\gamma^5\gamma^0 = \displaystyle{\slash \!\!\!\!p +
iM\over 2iM}$ \\\\ $v^{{\dagger}(\alpha)}(p)v^{(\beta)}(p) =
\displaystyle{\vert\vec p\vert\over M}$ \\
$v^{{\dagger}(\alpha)}(p)\gamma^0v^{(\beta)}(p) = 0$ \\
$v^{{\dagger}(\alpha)}(p)\gamma^5v^{(\beta)}(p) = \displaystyle{E\over
iM}\,i(\vec\sigma\!\cdot\!\hat p)_{\alpha\beta}$ \\
$v^{{\dagger}(\alpha)}(p)\gamma^5v^{(\beta)}(p)(-i\vec\sigma\!\cdot\!\hat
p)_{\beta\alpha'} = \displaystyle{E\over iM}\,\delta_{\alpha\alpha'}$ \\
$v^{{\dagger}(\alpha)}(p)\gamma^5\gamma^0v^{(\beta)}(p) =
-i(\vec\sigma\!\cdot\!\hat p)_{\alpha\beta}$ \\
$v^{{\dagger}(\alpha)}(p)\gamma^5\gamma^0v^{(\beta)}(p)(-i\vec\sigma\!\cdot\!\hat
p)_{\beta\alpha'} = -\delta_{\alpha\alpha'}$ \\
$v^{(\alpha)}(p)(-i\vec\sigma\!\cdot\!\hat p)_{\alpha\beta}\otimes
v^{{\dagger}(\beta)}(p)\gamma^5\gamma^0 = \displaystyle{\slash \!\!\!\!p -
iM\over 2iM}$ \\\\ $v^{{\dagger}(\alpha)}(E,-\vec
p)\gamma^5u^{(\beta)}(E,\vec p) = 0$ \\ $u^{{\dagger}(\alpha)}(E,-\vec
p)\gamma^5v^{(\beta)}(E,\vec p) = 0$ \\
$v^{{\dagger}(\alpha)}(p)\gamma^5\gamma^0u^{(\beta)}(p) = 0$ \\
$u^{{\dagger}(\alpha)}(p)\gamma^5\gamma^0v^{(\beta)}(p) = 0$
\\\\
\indent IV. `` Spin '' of the tachyon.
\\\\ The Lorentz transform generators leaving the tachyon equation invariant are :
$$\Sigma_{\mu\nu} = -{i\over4}
\left[\,\gamma_{\mu},\gamma_{\nu}\,\right]$$
Using the standard representation, one finds :
$$\Sigma_{0i} = -{i\over2}\pmatrix{0&\sigma_i\cr\sigma_i&0\cr},\ \ \ \ 
\ \ \ \ \Sigma_{ij} = -{1\over2}\,\varepsilon_{ijk}\pmatrix{\sigma_k&0\cr0&\sigma_k\cr}
$$ One then obtains the spin of the tachyon by looking at the Pauli--Lubanski operator :
$$W_{\mu} ={1\over 2}\,\varepsilon_{\mu\nu\rho\sigma}P^{\nu}J^{\rho\sigma}$$
with :
$$J_{\alpha\beta} = i\,(x_{\alpha}\partial_{\beta} - x_{\beta}\partial_{\alpha}) + \Sigma_{\alpha\beta}$$
One then has :$$W_{\mu} =
{1\over 2}\,\varepsilon_{\mu\nu\rho\sigma}P^{\nu}\Sigma^{\rho\sigma}$$
By choosing the characteristic vector : $P_{\mu} = ( 0, 0, 0, M )$, one obtains immediately:
$$W_{\mu}  = M (  \Sigma^{12}, \Sigma^{20},  \Sigma^{01},  0 )$$
The first three components of $W_{\mu}$ are nothing but the generators of the SU(1,1) Lie group.\\ And one finds :
\be W_{\mu}W^{\mu}   =   M^2   \left(    {\Sigma^{12}}^2   -   {\Sigma^{20}}^2   -
{\Sigma^{01}}^2\right)    =   {3\over   4}M^2\ee
From the relation $W_{\mu}W^{\mu} = -M^2s(s+1)$, one extracts : $ s = {\displaystyle -{1\over 2} \pm
{i\over\sqrt2}}\cdot$
\\\\
\indent V. Quantization.\\\\ We write the solution of (B.1) in the form :\beq{}&\displaystyle
\psi(x) =
\int\!\!{d^3p\over(2\pi)^3}{M\over E}\,\theta(\vert\vec p\vert-M)\sum_{\alpha,\beta=1}^2\Bigl[\,u^{(\alpha)}(p)(\vec\sigma\!\cdot\!\hat
p)_{\alpha\beta}\,b_{\beta}(\vec p)\,e\,^{\displaystyle
-ip\!\cdot\!x}\nonumber\\&\displaystyle+ v^{(\alpha)}(p)(\vec\sigma\!\cdot\!\hat
p)_{\alpha\beta}\,d^{\dagger}_{\beta}(\vec p)\,e\,^{\displaystyle
ip\!\cdot\!x}\,\Bigr]\eeq
The measure $\tilde dp =
\displaystyle{d^3p\over(2\pi)^3}{M\over E}\,\theta(\vert\vec p\vert-M)$ with $E =
\sqrt{\vec p\,^2 - M^2}$ is Lorentz invariant.\\ One quantizes
according to :$$\{ \,b_{\alpha}(\vec p) , b^{\dagger}_{\beta}(\vec p\,')\,\} =
(2\pi)^3{E\over M}\,\delta^3(\vec p-\vec p\,')\,(\vec\sigma\!\cdot\!\hat
p)_{\alpha\beta}$$$$\{ \,d_{\alpha}(\vec p) , d^{\dagger}_{\beta}(\vec p\,')\,\}
= (2\pi)^3{E\over M}\,\delta^3(\vec p-\vec p\,')\,(\vec\sigma\!\cdot\!\hat
p)_{\beta\alpha}$$All the other anticomutators vanish.  \\ One then
obtains :$$\{ \,\psi_i(\vec x, t), \pi_j(\vec y, t)\,\} = \{ \,\psi_i(\vec
x, t), \bigl(i\psi^{\dagger}(\vec y, t)\gamma^5\bigr)\!_j\,\} =
i\,\delta_{ij}\!\int\!\!{d^3p\over(2\pi)^3}\,\theta(\vert\vec p\vert-M)\,e\,^{\displaystyle i\vec p\!\cdot\!(\vec x-\vec y)} $$Or, using $(2.15)$ :  \be\{
\,\psi_i(\vec x, t), \pi_j(\vec y, t)\,\} =
i\,\delta_{ij}\,\hat\delta^{(3)}(\vec x - \vec y)\ee \\ The hamiltonian 
operator is then :$$H = \int\!\!d^3x\, {\cal H}(x) = \int\!\!\tilde
dp\,E\sum_{\alpha,\beta}\Bigl[\,b^{\dagger}_{\alpha}(\vec p)\,b_{\beta}(\vec p)-
d_{\alpha}(\vec p)\,d^{\dagger}_{\beta}(\vec p)\,\Bigr](\vec\sigma\!\cdot\!\hat
p)_{\alpha\beta}$$As a generalization, one has :$$P^{\mu} =
\int\!\!d^3x\,\psi^{\dagger}(x)\gamma^5\,i\partial^{\mu}\psi(x) = \int\!\!\tilde
dp \,p^{\mu}\sum_{\alpha,\beta}\Bigl[\,b^{\dagger}_{\alpha}(\vec
p)\,b_{\beta}(\vec p)- d_{\alpha}(\vec p)\,d^{\dagger}_{\beta}(\vec
p)\,\Bigr](\vec\sigma\!\cdot\!\hat p)_{\alpha\beta}$$In its normal form, it gives :
\be P^{\mu} = \int\!\!\tilde dp
\,p^{\mu}\sum_{\alpha,\beta}\Bigl[\,b^{\dagger}_{\alpha}(\vec p)\,b_{\beta}(\vec
p) + d^{\dagger}_{\beta}(\vec p)\,d_{\alpha}(\vec
p)\,\Bigr](\vec\sigma\!\cdot\!\hat p)_{\alpha\beta}\ee And, as
$(\vec\sigma\!\cdot\!\hat p)^{*}_{\alpha\beta} = (\vec\sigma\!\cdot\!\hat
p)_{\beta\alpha}$, one gets, as expected : $P^{\mu} = P^{\dagger\mu}$.
\\\\
 We obtain the usual comutation relations :$$\bigl[\,P^{\mu}, b_{\alpha}(\vec
p)\bigr] = -p^{\mu}b_{\alpha}(\vec p)\ ,\ \ \ \ \ \ \ \bigl[\,P^{\mu},
b^{\dagger}_{\alpha}(\vec p)\bigr] = p^{\mu}b^{\dagger}_{\alpha}(\vec
p)$$$$\bigl[\,P^{\mu}, d_{\alpha}(\vec p)\bigr] = -p^{\mu}d_{\alpha}(\vec p)\ ,\
\ \ \ \ \ \ \bigl[\,P^{\mu}, d^{\dagger}_{\alpha}(\vec p)\bigr] =
p^{\mu}d^{\dagger}_{\alpha}(\vec p)$$The charge operator is :$$Q =
\int\!\!d^3x\, j^0(x) = \int\!\!d^3x\, \psi^{\dagger}(x)\gamma^5\,\psi(x) =
\int\!\!\tilde dp \,\sum_{\alpha,\beta}\Bigl[\,b^{\dagger}_{\alpha}(\vec
p)\,b_{\beta}(\vec p) - d^{\dagger}_{\beta}(\vec p)\,d_{\alpha}(\vec
p)\,\Bigr](\vec\sigma\!\cdot\!\hat p)_{\alpha\beta}$$It follows that :
$$\bigl[\,Q, b_{\alpha}(\vec p)\bigr] = -b_{\alpha}(\vec p)\ ,\ \ \ \ \ \ \
\bigl[\,Q, b^{\dagger}_{\alpha}(\vec p)\bigr] = b^{\dagger}_{\alpha}(\vec
p)$$$$\hskip0.5cm\bigl[\,Q, d_{\alpha}(\vec p)\bigr] = d_{\alpha}(\vec p)\ ,\ \
\ \ \ \ \ \ \bigl[\,Q, d^{\dagger}_{\alpha}(\vec p)\bigr] = -d^{
\dagger}_{\alpha}(\vec p)$$
\\
\indent VI. Tachyonic states.
\\\\ We define :$$b_{\alpha}(\vec p)|\,0\!>\, = 0\ ,\ \ \ \ \ d_{\alpha}(\vec
p)|\,0\!>\, = 0$$$$b^{\dagger}_{\alpha}(\vec p)|\,0\!>\, = |\,\vec p\,,
\alpha\!>_T\ ,\ \ \ \ \ d^{\dagger}_{\alpha}(\vec p)|\,0\!>\, = |\,\vec p\,,
\alpha\!>_{\bar T}$$Then :\be<\!0\,|\,\psi(x)|\,\vec p\,, \alpha\!>_T\, =
u^{(\alpha)}(p)\,e\,^{\displaystyle -ip\!\cdot\!x}\ ,\ \ {}_{\bar T}\!<\!\vec
p\,, \alpha\,|\,\psi(x)|\,0\!>\, = v^{(\alpha)}(p)\,e\,^{\displaystyle
ip\!\cdot\!x}\ee\indent One finds the action of the quadri--momentum operator : 
$$P^{\mu}|\,\vec q\,, \gamma\!>_T\,
= \int\!\!\tilde dp \,p^{\mu}\sum_{\alpha,\beta}(\vec\sigma\!\cdot\!\hat
p)_{\alpha\beta}\Bigl[\,b^{\dagger}_{\alpha}(\vec p)\,b_{\beta}(\vec p) +
d^{\dagger}_{\beta}(\vec p)\,d_{\alpha}(\vec p)\,\Bigr]b^{\dagger}_{\gamma}(\vec
q)|\,0\!>$$$$= \int\!\!\tilde dp
\,p^{\mu}\sum_{\alpha,\beta}(\vec\sigma\!\cdot\!\hat
p)_{\alpha\beta}\,b^{\dagger}_{\alpha}(\vec p)(2\pi)^3{E\over M}\,\delta^3(\vec
p-\vec q)\,(\vec\sigma\!\cdot\!\hat p)_{\beta\gamma}|\,0\!>\,=
q^{\mu}\,b^{\dagger}_{\gamma}(\vec q)|\,0\!> = q^{\mu}|\,\vec q\,,
\gamma\!>_T$$The state vectors are normalized according to :  $$<\!0\,|\,0\!> \,= 1$$
$${}_T\!<\!\vec p\,, \alpha\,|\,\vec q\,,
\beta\!>_T \,= (2\pi)^3{E\over M}\,\delta^3(\vec p-\vec
q)\,(\vec\sigma\!\cdot\!\hat p)_{\alpha\beta}$$$${}_{\bar T}\!<\!\vec p\,,
\alpha\,|\,\vec q\,, \beta\!>_{\bar T}\, = (2\pi)^3{E\over M}\,\delta^3(\vec
p-\vec q)\,(\vec\sigma\!\cdot\!\hat p)_{\beta\alpha}$$

\end{document}